# Swift-heavy-ion-irradiation-induced enhancement in electrical conductivity of chemical solution deposited $La_{0.7}Ba_{0.3}MnO_3$ thin films

**R. N. Parmar, J. H. Markna, and D. G. Kuberkar** [a]
*Department of Physics, Saurashtra University, Rajkot – 360005, Gujarat, India*

**Ravi Kumar**
*Inter University Accelerator Centre, Aruna Asaf Ali Marg, New Delhi – 110067, India*

**D. S. Rana**
*Institute of Laser Engineering, Osaka University, 2-1 Yamadaoka, Suita, Osaka 565-0871, Japan*

**Vivas C. Bagve and S. K. Malik** [b]
*Tata Institute of Fundamental Research, Colaba, Mumbai – 400 005*

## *Abstract*

Epitaxial thin films of $La_{0.7}Ba_{0.3}MnO_3$ manganite, deposited using Chemical Solution Deposition technique, were irradiated by 200 MeV $Ag^{+15}$ ions with a maximum ion dose up to $1\times10^{12}$ ions/$cm^2$. Temperature- and magnetic field-dependent resistivity measurements on all the films (before and after irradiation) reveal a sustained decrease in resistivity with increasing ion dose. A maximum dose of $1\times10^{12}$ ions/$cm^2$ suppresses resistivity by factors of 3 and 10, at 330 K [insulator-metal (I-M) transition] and at 10 K, respectively. On the other hand, with increasing ion dose, the magnetoresistance (MR) enhances in the vicinity of I-M transition but decreases at low temperatures. These results, corroborated by surface morphology of films, suggest that the origin of such properties lies in the irradiation induced improved crystallinity and epitaxial orientation, enhanced connectivity between grains, and conglomeration of grains which result in better conductivity at grain boundaries.

[a] Email: dgkuberkar@rediffmail.com

[b] Present address: International Centre for Condensed Matter Physics- ICCMP, University of Brasilia, Brasilia, Brazil

Optimal divalent cation-doped manganites of the general formula $La_{0.7}A_{0.3}MnO_3$ (A = $Ca^{2+}$, $Sr^{2+}$, $Pb^{2+}$, $Ba^{2+}$, etc.), have attracted keen interest of the scientific community owing to the application potential of colossal magnetoresistance (CMR) exhibited by them. This CMR effect is dominant in the vicinity of concurrent insulator-metal transition temperature ($T_P$) and ferromagnetic Curie temperature ($T_C$). The metallic conductivity in ferromagnetic region of these manganites occurs due to Zener-double exchange (ZDE) mechanism between $Mn^{3+}$ ($3d^4$) and $Mn^{4+}$ ($3d^3$) ions via oxygen-2p orbitals. This further depends upon two major factors, namely, the tolerance factor and the $Mn^{3+}/Mn^{4+}$ ratio [for a review, see ref. 1]. Also, it is well established that electronic transport in these manganites is extremely sensitive to various kinds of disorders such as A-site cation disorder [2,3], grain boundary effects [4,5], external or internal pressure [6], ion irradiation [7-10], etc..

Swift Heavy Ion (SHI) irradiation is an efficient tool for creation of point defects, vacancies, columnar defects and localized strain in various materials (mainly thin films) which in turn affect their crystallographic, surface and physical properties [7-11]. SHI irradiation on manganite thin films mostly leads to detrimental effects on structural and transport properties. For instance, the irradiation results in enhancement of resistivity, suppression of $T_p$, etc [7, 12]. Though the deteriorated properties of the irradiated films have adverse effect with respect to their applications as data storage devices, their application possibilities improve as uncooled bolometers, magnetic field sensors, etc. [13]. However, all such endeavors to control the physical properties of manganites by irradiation have been made on the thin films synthesized either using Pulsed Laser Deposition (PLD) or other physical methods but there have been no attempts to study such irradiation effects in manganite films synthesized by chemical methods. In this context, we have synthesized $La_{0.7}Ba_{0.3}MnO_3$ (LBMO) thin films by chemical solution deposition (CSD) method. In this letter, we report the effect of SHI irradiation on the physical properties of LBMO thin films synthesized by CSD method. Our findings are unique and interesting in the sense that the irradiation effects on electronic transport and magnetoresistance (MR) of CSD grown LBMO thin films are totally opposite to those reported for PLD grown manganite thin films [9-12]. The selection of highly energetic

200 MeV $Ag^{+15}$ ions has been made in the present irradiation studies, since these ions can easily pass through the entire thickness of the thin film and create columnar tracks before striking the substrates. This results in a uniform distribution of strain in the films which is not the case with low energy ion implantation [12].

All the $La_{0.7}Ba_{0.3}MnO_3$ thin films were grown on single crystalline $LaAlO_3$ *(h00)* substrate using Chemical Solution Deposition (CSD) technique which is relatively simple, cost effective and requires only low synthesis temperature as compared to PLD or MBE techniques. The mixing, stirring and heating of appropriate stoichiometric quantities of the metal acetates in distilled water and acetic acid resulted in a clear solution of the constituents used for deposition. Deposition of thin films using automated spin coater was then followed by heating and annealing at 1000°C in an oxygen environment. The films were characterized by X-ray diffraction (XRD) and AFM measurements for their structural and microstructural properties. Using d.c. four probe resistivity technique, the electrical resistivity of all the films was measured i) as a function of temperature in the range of 5 -385 K in zero-field and in an applied field of 9 T and ii) as a function of magnetic field (up to 9 Tesla) at various temperatures [PPMS, Quantum Design]. SHI irradiation was performed at Inter University Accelerator Centre (IUAC), New Delhi, India, using a 15 UD tandem accelerator. 200 MeV $Ag^{+15}$ ion beam [14] with a beam current $^{15}$ ~ 0.3 pnA [15] was used with ion doses of $7.5\times10^{10}$ and $1\times10^{12}$ ions/$cm^2$.

The XRD patterns (fig.1) of pristine and irradiated $La_{0.7}Ba_{0.3}MnO_3$ thin films reveal that these films are *(0 0 l)* orientated on single crystal LAO *(h 0 0)* substrate. A small amount (of ~ 4-5 %) of BaO secondary phase was also detected (marked by an asterisk in fig. 1). The values of out-of-plane lattice parameter, calculated from the XRD patterns, was found to increase from 3.9091(3) Å for the pristine film to 3.9051(3) Å and 3.9140(3) Å, respectively, for $7.5\times10^{10}$ ions/$cm^2$ and for $1\times10^{12}$ ions/$cm^2$ irradiated films. The value of residual microstress in the films due to the substrate-material mismatch have been calculated using the formula, $\delta\% = \frac{d_{substrate} - d_{thin\,film}}{d_{substrate}} \times 100$, which is ~ -2.88 % for pristine and ~ -2.7 % for the film irradiated with $1 \times 10^{12}$ ions/$cm^2$. This suggests a decrease in compressive strain in the films after irradiation. Furthermore, the full width at half maximum (FWHM) of *(0 0 1)* and *(0 0 2)* peaks of the LBMO decreases with increasing irradiation. For instance, the FWHM of *(0 0 2)* peak decreases from 0.224º for pristine film to 0.205º for $1\times10^{12}$ ion/cm$^2$ irradiated films. This signifies an improvement in the crystalline structure after the irradiation.

Figure 2 shows resistivity (ρ) vs. temperature (T) plots for pristine and irradiated $La_{0.7}Ba_{0.3}MnO_3$ films in zero-field and in a field of 9 T. The peak resistivity of pristine LBMO at insulator-metal transition temperature $T_p$ (330 K) is ~ 110 (mΩ cm), which is in good agreement with the reported value for PLD deposited LBMO thin film [16]. Interestingly, the irradiation induces a significantly consistent suppression in resistivity over the whole temperature range investigated. For instance, the resistivity decreases by a factor of 2 and 4, respectively, at 330 K and 10 K with an ion dose of $7.5\times10^{10}$ ions/cm$^2$, while it decreases by a factor of 3 and 10, respectively, at 330 K and 10 K with an ion dose of $1\times10^{12}$ ions/cm$^2$. Also, it is interesting to note that, with increasing irradiation, $T_p$ remains nearly unaffected as it shifts slightly from 330 K for pristine to 337 K for $1\times10^{12}$ ions/cm$^2$ irradiated film. This kind of irradiation induced suppression of resistivity, with $T_p$ remaining unaffected in CSD films, is a unique observation vis-à-vis the earlier reported irradiation induced enhancement of resistivity and suppression of $T_p$ in PLD grown films.

To obtain an insight on the origin of these effects, we have plotted magnetoresistance (MR%) versus magnetic field (H) isotherms at various temperatures for all the $La_{0.7}Ba_{0.3}MnO_3$ films [fig. 3]. It is seen that all the films exhibit maximum MR in the vicinity of $T_p$ which decreases with decreasing temperature. We discuss these MR results as a function of temperature as follows.

i) At a low temperature of 10 K, the pristine film exhibit low field MR (LF-MR) ~ 5 % (in a field of < 1 T). This uncharacteristic MR behavior of an oriented thin film partially resembles the LF-MR behavior of polycrystalline samples, the origin of which is attributed to the intergrain spin polarized tunneling (SPT) of charge carriers that depends upon the alignment of the magnetic moment of the ferromagnetic domains [17]. It is interesting to note that the LF-MR for the irradiated films at 10K decreases to ~ 2-3 % and to zero for respective ion doses of $7.5\times10^{10}$ ions/cm$^2$ and $1\times10^{12}$ ions/cm$^2$. Also, the high field MR [HF-MR] (in a field >1 Tesla) at 10 K decreases from ~ 15 % for the pristine film to nearly zero for the film irradiated with $1\times10^{12}$ ions/cm$^2$. This points towards irradiation induced enhancement in connectivity between the grains, reduction in pinning of Mn-ion spins at grain boundaries, etc.

ii) At an intermediate temperature of 240 K, the pristine film shows LF-MR < 3 % and irradiated films show nearly zero. The HF-MR is nearly 30 % for pristine film which reduces to 20 % and 10 %, respectively, for the films irradiated with $7.5\times10^{10}$ ions/cm$^2$ and $1\times10^{12}$ ions/cm$^2$.

iii) In the vicinity of $T_P$ at 330 K, the LF-MR of ~ 8 %, 12 % and 18 % is observed for the pristine, the $7.5\times10^{10}$ and $1\times10^{12}$ ions/cm$^2$ irradiated thin films, respectively. The HF-MR observed for the pristine film is ~ 48 %, while successive irradiation doses of $7.5\times10^{10}$ and $1\times10^{12}$ ions/cm$^2$, enhance the HF-MR to ~ 55 % and 60 %, respectively. Such an irradiation induced enhancement in MR at $T_p$ may be attributed to the field induced delocalization of charge carriers due to the structural disorder at grain boundaries and at Mn-O-Mn bond angles [17,18].

The above-mentioned results, depicting an irradiation induced suppression in MR at low temperatures and an enhancement of MR at high temperatures around $T_p$, underline the importance of deteriorating grain boundary contribution to MR. Moreover, the irradiation induced suppression of LF-MR indicates further the improvement in crystallographic orientation and parallel alignment of magnetic domains. This inference is supported by increasing line intensity in XRD patterns [fig. 1]. At low temperatures, HF-MR arises due to poor connectivity between grains, pinning of Mn spin at grains boundaries, etc. [18]. The suppression of HF-MR at 10 K indicates irradiation induced enhancement in grain connectivity. Such granular contribution is also evident from the MR behavior at $T_p$; the enhancement in MR with increasing ion dose at 330 K implies that the magnetic order at the grain boundaries improves, thus, allowing a better electronic transport.

To further probe the granular contribution, we have recorded AFM micrographs for the pristine and the irradiated La$_{0.7}$Ba$_{0.3}$MnO$_3$ films, and the results are shown in fig. 4. The microstructures of all the films show an island type of growth which is mostly responsible for the non-uniform distribution of strain in the film [6]. The observed surface roughness for the pristine film is ~ 12.35 % which reduces to 9.25 % for $7.5\times10^{10}$

ion/cm$^2$ and to 8 % for 1×10$^{12}$ ion/cm$^2$ irradiated films. The values of average grain size obtained from AFM pictures are 50 nm, 80 nm and 70 nm for the pristine film and for 7.5×10$^{10}$ and 1×10$^{12}$ ions/cm$^2$ irradiated films, respectively. Apart from a moderate reduction in surface roughness which points towards better orientation, a significant enhancement in grain size with increasing irradiation suggests a dominant possibility of conglomeration of two or more grains to form a bigger size grain. This would result in a decreased grain boundary region and, consequently, an improved electronic transport at the grain boundaries which accounts for irradiation induced reduction of resistivity in CSD grown thin films. To evaluate the effect of irradiation on the application aspects of these CSD films, we have calculated the temperature and the field sensitivity of resistivity. These parameters, quantified as the temperature coefficient of resistance [TCR % = (1/R×dR/dT)×100] and the field coefficient of resistance [FCR % = (1/R×dR/dH)×100], are plotted for La$_{0.7}$Ba$_{0.3}$MnO$_3$ films in the insets of fig. 2 & fig. 3, respectively. In irradiated films, a significant positive enhancement in TCR values and enhancement in negative values of FCR near T$_P$ (slightly above room temperature) suggests the usefulness of ion irradiated CSD films for bolometric and magnetic sensor applications.

To sum up, we find that the effects of swift ion irradiation on microstructural, electrical and magnetotransport properties of La$_{0.7}$Ba$_{0.3}$MnO$_3$ manganite thin films grown by CSD method are opposite to the irradiation effects on manganite films grown by PLD. The SHI irradiation induced decreased resistivity (consistent with improved crystallinity, microstructure and variations in magnetoresistance of La$_{0.7}$Ba$_{0.3}$MnO$_3$ thin films) opens up new avenue for engineering electronic and magnetotransport properties of manganites for applications employing swift heavy ion irradiation.

DGK and RNP are thankful to IUAC, New Delhi, for financial assistance in the form of UFUP project for SHI irradiation experiment. The experimental help by Prof. A. K. Raychaudhuri and Dr.(Ms.) B. Ghosh in the use of CSD method and of Dr. R. J. Choudhary in irradiation experiment is gratefully acknowledged.

14. To generate 200 MeV Ag ion beam using pelletron accelerator, first negatively charged Ag ions are produced and accelerated with the chosen terminal potential (V). To further accelerate the ions, charge stripping of the ions is performed to convert the Ag ions to higher charge state, e.g. +15, by passing the ion through a stripper foil (carbon foil). This high charge state ion is further accelerated by the same potential to carry higher energy. Charge state q of the ion depends purely on the requirement of the energy of the beam and can be calculated using the formula, $E = (1+q)V$. With E=200Mev, V~ 12.5 MV, the value of q is 15. The charge state is merely used to achieve the desired energy of 200 MeV energy

15. pnA stand for particle nanoampere current. In the case of ion irradiation, the unit for the current is particle Ampere. In the present experiment, the beam current is ~ 4.5 nA. To calculate pnA, the current is divided by the charge state which means that each particle/ion carries electronic charge equivalent to its charge state.

Figure 1:

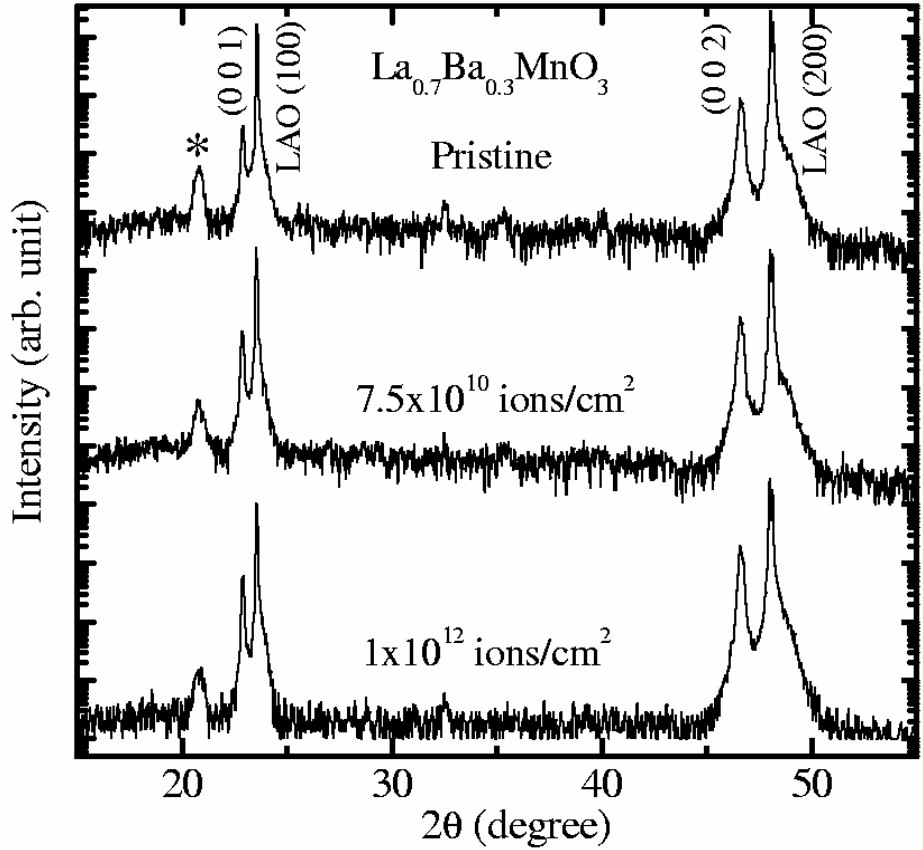

Figure 1: XRD patterns of pristine and irradiated $La_{0.7}Ba_{0.3}MnO_3$ thin films. The asterisk (*) denotes a peak from BaO impurity phase.

Figure 2:

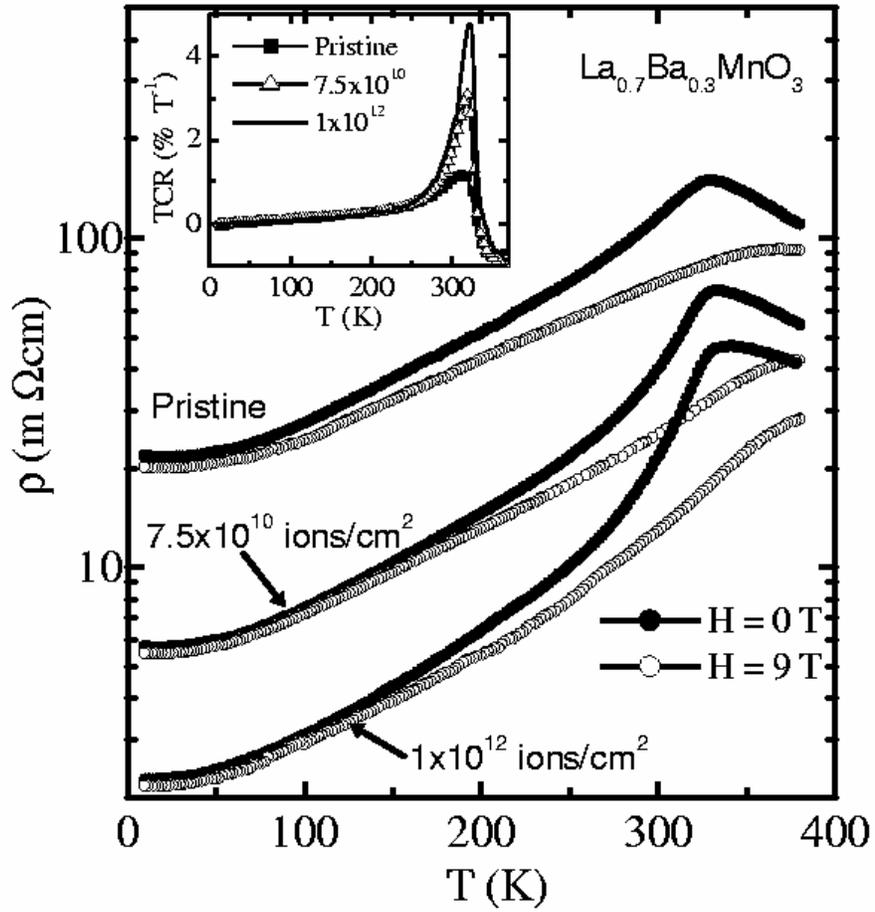

Figure 2: Resistivity [ρ] vs. temperature [T] plots for pristine and irradiated $La_{0.7}Ba_{0.3}MnO_3$ thin films. Inset shows temperature co-efficient of resistance [TCR] vs. temperature [T] plots for pristine and irradiated thin films.

Figure 3:

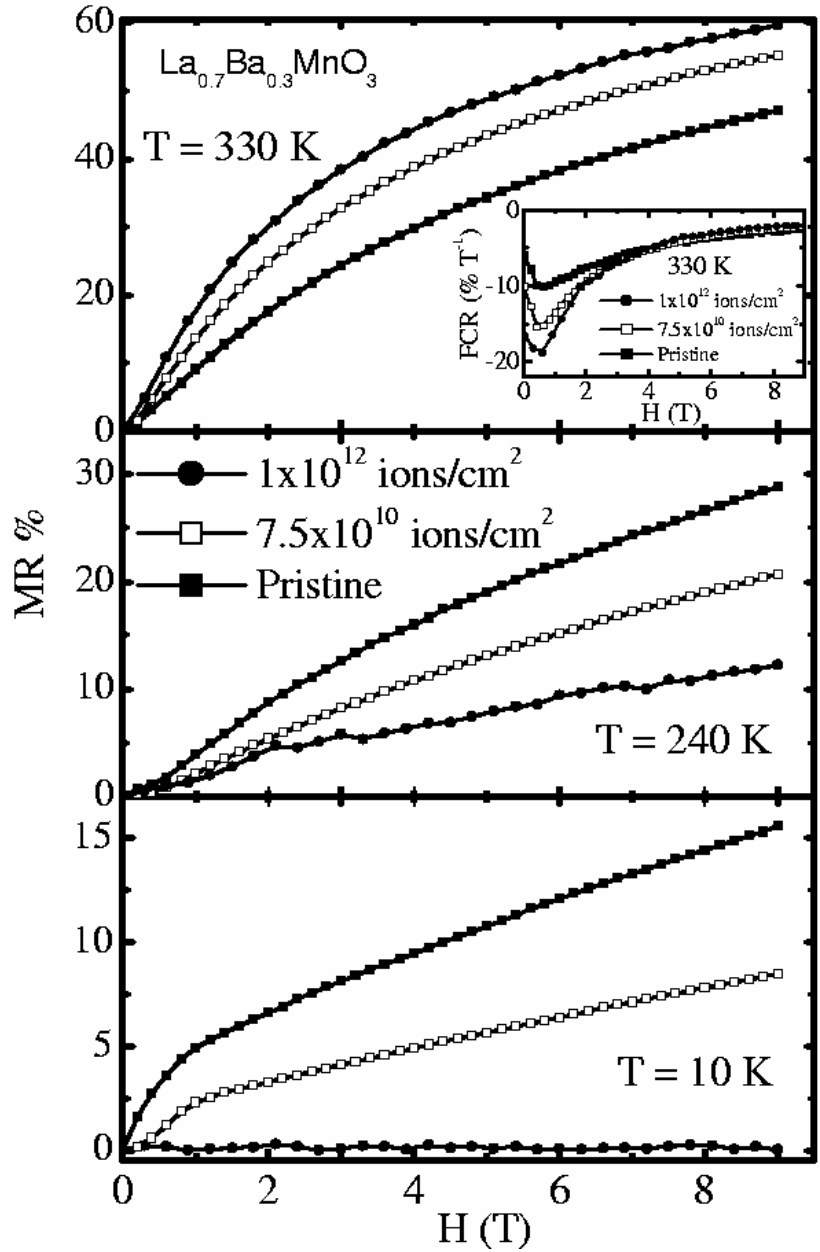

Figure 3: Magnetoresistance [MR (%)] vs. magnetic field [H] isotherms for pristine and irradiated $La_{0.7}Ba_{0.3}MnO_3$ thin films. Inset shows field co-efficient [FCR] vs. magnetic field [H] plots at 330 K for pristine and irradiated thin films.

Figure 4:

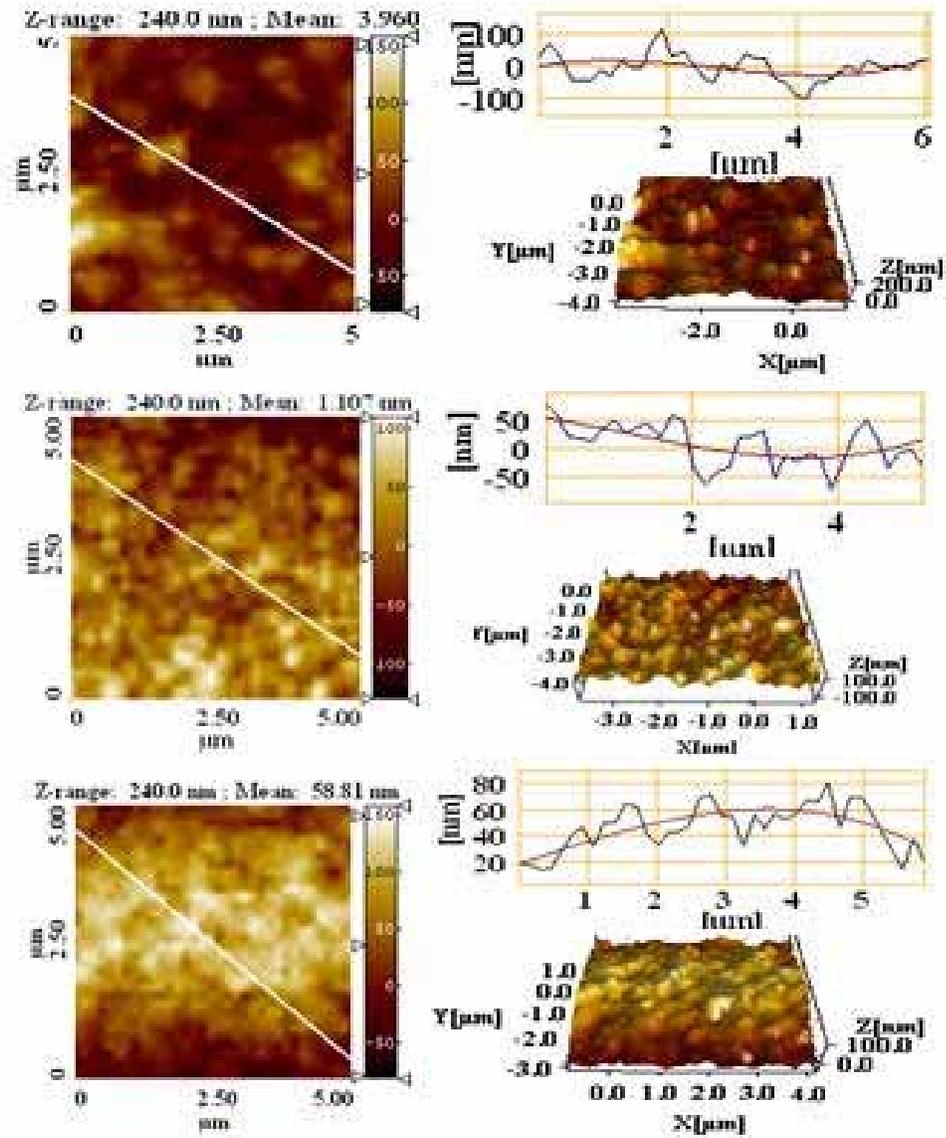

Figure 4: AFM micrographs of (a) pristine, (b) $7.5\times10^{10}$ ions/cm$^2$ irradiated and (c) $1\times10^{12}$ ions/cm$^2$ irradiated La$_{0.7}$Ba$_{0.3}$MnO$_3$ thin films along with their respective surface roughness histograms.